\documentclass[11pt,a4paper]{article} 
\pdfoutput=1
\usepackage{jcappub}

\usepackage{graphicx,wasysym}
 \pdfoutput=1

\def\K{K{\"a}hler}
\newcommand{\ft}[2]{{\textstyle\frac{#1}{#2}}}
\def\rmi{{\rm i}}

\def\rme{{\rm e}}

\newcommand{\mr}{\mathcal{R}}

\newcommand{\rf}[1]{(\ref{#1})}

\def\rmi{{\rm i}}

\def\rme{{\rm e}}

\def\be{\begin{equation}}
\def\ee{\end{equation}}
\def\bea{\begin{eqnarray}}
\def\eea{\end{eqnarray}}
\def\K{K{\"a}hler}

\title{\rm {\bf \huge Superconformal Generalization of the Chaotic Inflation Model ${\lambda\over 4} \phi^4 - {\xi\over 2}\phi^2 R$}}
\author{\rm \Large {\bf Renata Kallosh and Andrei Linde}}

\affiliation{Stanford Institute for Theoretical Physics and Department of Physics, \\ Stanford University, Stanford, CA 94305 USA}
 \emailAdd{kallosh@stanford.edu}\emailAdd{alinde@stanford.edu} 
\abstract{A model of chaotic inflation based on the theory of a scalar field with potential $\lambda\phi^4$ perfectly matches the observational data if one adds to it a tiny non-minimal coupling to gravity $-{\xi\over 2}\phi^{2} R$ with $\xi \gtrsim 0.002$. We describe embedding of this model  into the superconformal theory with spontaneous breaking of superconformal symmetry, and into supergravity. A model with small  $\xi$ is technically natural: setting the small  parameter $\xi$ to zero leads to a  point of   enhanced symmetry in the underlying superconformal theory.

}

\begin{document}
\maketitle

\section{Introduction}

The simple chaotic inflation model with potential $\lambda\phi^{4}$ in the Einstein frame \cite{Linde:1983gd} is conclusively ruled out by the data due to the high level of the tensor-to-scalar ratio $r$ which it predicts. Meanwhile the same model $\lambda \phi^4$, which includes the non-minimal gravitational coupling $\xi \phi^2 R/2$  in the Jordan frame, makes a dramatic comeback and is in perfect agreement with the Planck2013  data \cite{Ade:2013rta} for $\xi/2 \gtrsim  10^{{-3}}$  \cite{Okada:2010jf,Bezrukov:2013fca}, see Fig. 1.

Effects of non-minimal coupling to gravity on inflation has received a lot of attention over the years \cite{Futamase:1987ua,Salopek:1988qh,Makino:1991sg,Sha-1,Einhorn:2009bh,Ferrara:2010yw,Lee:2010hj,Ferrara:2010in}.  
The recent revival of interest to these models was related to the possibility that
the Higgs field of the standard model may play the role of the inflaton \cite{Salopek:1988qh,Sha-1,Einhorn:2009bh,Ferrara:2010yw,Lee:2010hj,Ferrara:2010in}, which would require $\lambda = O(1)$ and  $\xi \gg 1$. However, the inflaton does not have to be a Higgs field, its quartic coupling $\lambda$ is not constrained by the standard model phenomenology. Therefore $\lambda$ can be small, which means in turn that the non-minimal coupling $\xi$ does not have to be large. It is amazing that adding to the Lagrangian the term $\xi \phi^2 R/2$ with a minuscule coefficient $\xi/2 >  10^{{-3}}$ is sufficient to make the simple chaotic inflation model $\lambda \phi^4$ viable.
Thus one may wonder whether there is something special about the model $\lambda\phi^{4}$ with nonminimal coupling to gravity.
\begin{figure}[ht!]
\centering
\vskip 0.2cm \includegraphics[scale=0.6]{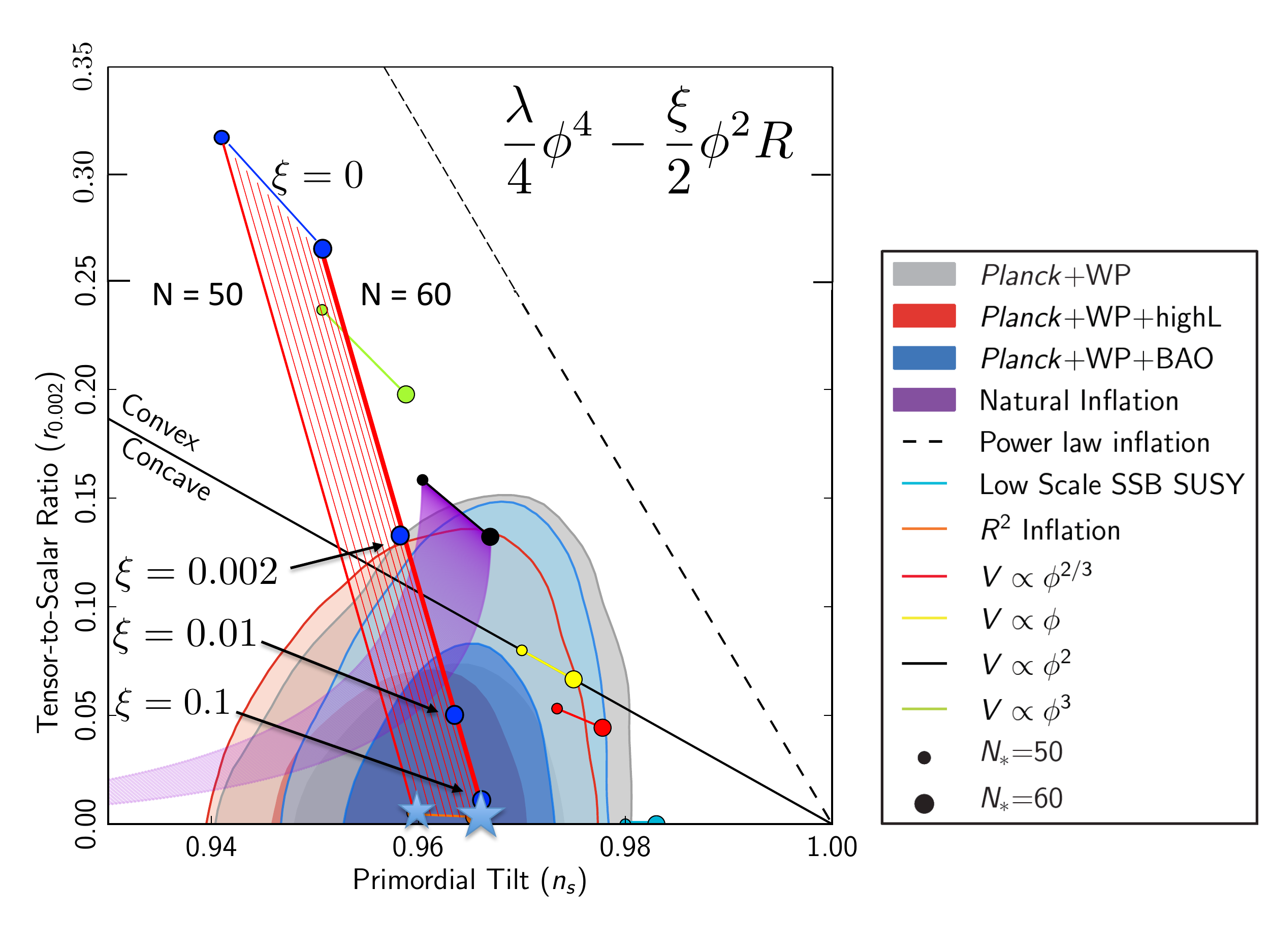}
\caption{Red stripes show inflation parameters $n_{s}$ and $r$ for the model $\lambda\phi^{4}/4$ for different values of $\xi$ in the  term $\xi\phi^{2} R/2$. The top range corresponds to $\xi = 0$. Different stripes correspond to different number of e-foldings $N$. The right one corresponds to $N = 60$, the left one corresponds to $N = 50$.}
\label{potential}
\end{figure}

The model $\lambda \phi^4/4 - \xi \phi^2 R/2$ with conformal coupling $\xi = -1/6$ was often studied in the cosmological literature of the 70's, because of its conformal invariance, which simplified investigation of classical and quantum evolution of scalar fields in the conformally flat Friedmann space. However, with invention of inflationary theory this model was practically abandoned because it did not lead to inflation. A related problem was also found in the context of the  string theory KLMMT model of inflation \cite{Kachru:2003sx} where the conformal coupling of the D3 brane makes the realization of the slow-roll inflation more complicated:  the D3-brane in a highly warped region  is dual to an almost conformal four-dimensional field theory  and the scalar field describing the motion of the brane is conformally coupled.

The theory $\lambda \phi^4/4 - \xi \phi^2 R/2$ with $\xi \geq 0$ leads to inflation \cite{Linde:1983gd,Futamase:1987ua,Salopek:1988qh,Makino:1991sg,Sha-1}, but the original conformal invariance of the theory $\xi = -1/6$ is now broken explicitly by the term $\xi \phi^2 R/2$. 
Moreover, the theory $\lambda \phi^4/4$ with $\xi = 0$ is ruled out by observations, and until recently it was not easy to implement models with $\xi \not = 0$ in the context of supergravity. In the standard textbook formulation of supergravity, all scalars automatically have minimal coupling to gravity, $\xi = 0$. This may have implied that the models with non-minimal coupling to gravity are incompatible with supersymmetry. If correct, it would be a strong argument against cosmological models with non-minimal coupling.

A solution of this problem was found few years ago \cite{Einhorn:2009bh,Ferrara:2010yw,Lee:2010hj,Ferrara:2010in}. In particular, in collaboration with Ferrara, Marrani and Van Proeyen, we reformulated supergravity in a way that provides a better link to its hidden superconformal symmetry \cite{Ferrara:2010yw,Ferrara:2010in}. This allowed us to introduce non-minimal coupling of scalars to gravity and to implement Higgs inflation scenario in supergravity. 

The presentation of our results in  \cite{Ferrara:2010yw,Ferrara:2010in} required 
extensive use of advanced supergravity formalism. In addition, if one wants to force the standard Higgs boson to play the role of the inflaton, as suggested in \cite{Sha-1}, one can do it only in rather advanced phenomenological models such as NMSSM. These complexities could obscure basic qualitative ideas of our approach and physical motivation of some of the cosmological models which we developed. Therefore in this paper we will take a step back, relax the requirement relating the inflaton field to the Higgs field, and describe our method and the resulting models in a more intuitive way. 

We will focus here on the  identification of symmetries of superconformal theory, some of which become flat directions of the supergravity potential for $\xi = 0$.  This feature of the superconformal models has been already used in practical inflationary model building in  \cite{Kallosh:2010ug,Linde:2011nh}. In this paper the issue of enhanced symmetries will be studied in the context of the critical points of the parameter $\Delta$, the deformation parameter of  the canonical superconformal model (CSS) described in Appendix A.

Our starting point is an old observation that the model with a single {\it real} conformal scalar (conformon) with the Lagrangian 
$\sqrt{-{g}}\left[{1\over 2}\partial_{\mu}\chi \partial^{\mu}\chi  +{ \chi^2\over 12}  R({g})\right]$
is equivalent to pure Einstein theory. Indeed, this theory is invariant with respect to conformal transformations $g'_{\mu\nu} = \rme^{-2\sigma(x)} g_{\mu\nu}$,
$\chi' =  \rme^{\sigma(x)} \chi$. Therefore without any loss of generality one can simply take $\chi = \sqrt{6}M_P$. One may regard it as a spontaneous breaking of conformal symmetry, which reduces this theory to the 
standard Einstein theory of gravity.

Following  \cite{Ferrara:2010in}, we will begin Section \ref{real} with a description of a closely related conformally invariant model of two real scalar fields, the inflaton $\phi$ with the potential $\lambda\phi^{4}$, and the conformon $\chi$. Just as in the example given above, the model will be invariant under local conformal transformations, but this invariance will be {\it spontaneously} broken if one considers a constant solution of equations of motion for the conformon field, $\chi = \sqrt{6}M_P$. We will explain that this simple model does not describe inflation unless we change the value of the parameter  $\xi = -1/6$. It is hard to do so in the model of a single real inflaton field $\phi$ without breaking conformal invariance explicitly. 

Fortunately, one can do it in a theory of a  {\it complex} scalar field $\phi$ and a complex conformon\footnote{These complex fields are coordinates of a \K\, geometry, which means that the kinetic terms for these scalars are defined by a \K\, potential.} as we will show in Section \ref{scmatter}. In this case, the parameter $\xi$ corresponding to the non-minimal coupling of the inflaton multiplet to gravity still has the conformal value $\xi=-1/6$ in the coupling of $\phi\bar \phi$ to the curvature. However, one can add  terms with coupling $\Delta$ of the following structure
\be
-{1\over 6} \,  \phi \bar \phi \, R +{\Delta\over 2} (\phi^2+\bar \phi^2)  R -\partial \phi \partial\bar \phi\ .
\ee
This means that the parameter $\xi$ takes two different values for the real and imaginary parts of the  field $\phi$  \cite{Ferrara:2010in}: 
\be
\xi_{\rm re} = -{1\over 6} + \Delta\, ,  \qquad  \xi_{\rm im} = -{1\over 6} - \Delta \ .
\ee
By taking $\Delta = \pm 1/6$, one can make $\xi = 0$ either for the real part of the filed $\phi$, or for the imaginary one. For  $\Delta=1/6$
\be
\xi_{\rm re}=0\, , \qquad \xi_{\rm im}= -{1\over 3}
\ee
and vice versa for $\Delta=-1/6$, \, $\xi_{\rm re}=-1/3\, ,  \, \xi_{\rm im}= 0$.

 By increasing $|\Delta|$ just $2\cdot 10^{-3}$ beyond its critical value $|\Delta|_{\rm cr} =1/6$, one can reproduce the values of the coupling $\xi \geq 0.002$ required for the successful inflationary scenario compatible with Planck observations, see Fig. 1. {\it This modification, supplemented by a proper modification of kinetic terms in the action, leaves the underlying action conformally invariant for an arbitrary $\Delta$}.  Once again, this invariance is spontaneously broken by the conformon field.
Local conformal invariance of the model with the non-vanishing $\Delta$ studied in Section \ref{scmatter} is quite non-trivial; it is possible only if scalars, including a set of physical scalars as well as an extra conformon field, form  an embedding \K\, geometry where the \K\, potential has a particular Weyl  weight.

We were able to construct this model  by making a certain simplification of our investigation  based on supergravity. In fact, it was supergravity that provided  the original motivation for this work. Indeed, conformal invariance plays an important role in the formulation of supergravity. It was known \cite{Freedman:1976hn} how to make supergravity locally scale invariant from the early days of this theory.
The full compensating supermultiplet was first  added to  supergravity in \cite{Siegel:1977hn}, in superfields, and in \cite{Kaku:1978ea}, in components,  to promote its symmetry to the superconformal one. This added not only local scale invariance but also other local symmetries: $U(1)\, \mr$ symmetry,  special conformal symmetry and  special supersymmetry.

The general formulation of $N = 1$ supergravity often  starts with the superconformal theory \cite{Siegel:1977hn}-\cite{Kallosh:2000ve}. 
If in a superconformal theory with $-{1\over 6} {\cal N}(X, \bar X) R$ coupling to gravity  \cite{Kallosh:2000ve}, one makes a choice that the \K\, potential  of the embedding space is constant, ${\cal N}(X, \bar X)=-3$, 
one derives the standard textbook formulation of supergravity with ${1\over 2} R$ in the action (minimal coupling),
\cite{Siegel:1977hn}-\cite{Weinberg:2000cr}.
 Once this step is made, the theory is formulated in the Einstein frame, all scalars have minimal coupling to gravity, and the superconformal origin of supergravity becomes well hidden. The new supergravity textbook \cite{Freedman:2012zz} offers a detailed superconformal derivation of supergravity.

In \cite{Ferrara:2010yw} we performed an alternative gauge-fixing  of the superconformal theory,
which allows to derive the supergravity action in an arbitrary Jordan frame, with physical scalars coupled to gravity in the form $\Phi(z, \bar z) R$. This provided a complete locally supersymmetric theory for scalars with a  nonminimal coupling to gravity. In this new framework, we identified in \cite{Ferrara:2010in} a particularly simple class of supergravity models,  canonical superconformal supergravity (CSS) models, see Appendix A in this paper. In these models, with local supersymmetry, {\it the kinetic terms in the Jordan frame are canonical, and the scalar potential is the same as in the global theory}.  The pure supergravity part of the total action has a local Poincar{\'e} supersymmetry, whereas the chiral and vector multiplets coupled to supergravity have a larger local superconformal symmetry.  

In this context, inflation in the theory $\lambda\phi^{4}$ with nonminimal coupling to gravity appears in the most natural way, as a single parameter deformation of  CSS, defined by a parameter $\Delta$. The superconformal version of the ${\lambda\over 4} \phi^4 - {\xi\over 2}\phi^2 R$ is presented in Section \ref{SUGRA}. We  discuss the phenomenology of our model in Section 5.

Let us reiterate the main reason why we developed these models instead of simply adding a term ${\xi\over 2}\phi^{2} R$ to the theory: We wanted to implement inflation in supergravity, and we did not want to break conformal symmetry of the theory explicitly.  One of our goals was to preserve conformal symmetry, and break it spontaneously, in the least invasive way by $\Delta$ deforming a CSS model, when the conformon field acquires  a nonzero vacuum expectation value.

\section{Conformally invariant model for real scalars} \label{real}

Following \cite{Ferrara:2010in}, we will begin our discussion with a simple  conformally invariant toy model of gravity and two real scalar fields, $\phi$ and $\chi$:
\begin{equation}
\mathcal{L}_{\rm toy} = \sqrt{-{g}}\left[{1\over 2}\partial_{\mu}\chi \partial^{\mu}\chi  +{ \chi^2\over 12}  R({g})- {1\over 2}\partial_{\mu} \phi\partial^{\mu} \phi   -{\phi^2\over 12}  R({g}) -{\lambda\over 4}\phi^4\right]\,.
\label{toy}
\end{equation}
The field $\chi(x)$ is referred to as  a conformal compensator, which we will call `conformon'  \cite{Kallosh:2000ve}. This
theory is locally conformal invariant under the following
transformations: 
\be g'_{\mu\nu} = \rme^{-2\sigma(x)} g_{\mu\nu}\,
,\qquad \chi' =  \rme^{\sigma(x)} \chi\, ,\qquad \phi' =  \rme^{\sigma(x)}
\phi\ . \label{conf}\ee 
Note that the kinetic term of the conformon
 $\chi$ has a wrong sign. This is not a problem because there
are no physical degrees of freedom associated with it; the field $\chi$
can be removed from the theory by fixing the gauge symmetry
(\ref{conf}). If we choose the gauge $\chi(x) = \sqrt{6}M_P$, the
$\phi$-terms in (\ref{toy}) reduce to the Einstein action. The full
Lagrangian in the Jordan frame is
\begin{equation}
\mathcal{L}_{\rm total }= \mathcal{L}_{\rm E } + \mathcal{L}_{\rm conf }= \sqrt{-{g}}\, M_P^2{  R({g})\over 2}-  \sqrt{-{g}}\left[{1\over 2}\partial_{\mu} \phi \partial^{\mu} \phi   +{ \phi^2\over 12} R({g}) +{\lambda\over 4}\phi^4\right]\,.
\label{toy2}
\end{equation}
It consists of two parts, the Einstein Lagrangian $
\sqrt{-{g}}\,M_P^2{ R({g})\over 2} $, which is not conformally
invariant, and the conformally invariant theory of the canonically
normalized scalar field $\phi$,
\be
 \mathcal{L}_{\rm conf }=-  \sqrt{-{g}}\left[{1\over 2}\partial_{\mu} \phi \partial^{\mu} \phi \,   +{ \phi^2\over 12} R({g}) +{\lambda\over
 4}\phi^4\right]\,.
\label{c}
\ee

As we already mentioned, theories of this type played a very important
role in the development of particle physics and cosmology many decades
ago.  One of the main reasons is that the Friedmann
universe is conformally flat. By making a conformal transformation, one
could represent equations of motion of the scalar field in the Friedmann
universe in terms of equations of motion of a conformally transformed
field in Minkowski space. This is a tremendous simplification for 
investigation of classical and quantum processes in an expanding universe.

The theory (\ref{toy}) is unique if we require that the local conformal
symmetry of the $\phi$ part of the action, which has canonical kinetic
terms, should be preserved after the gauge fixing. It is determined by
the condition that the conformon $\chi(x)$ is decoupled from
the inflaton field  $\phi(x)$.  The trace of the energy-momentum tensor of the $\phi$ field vanishes on shell and there is no inflation. If we would change the factor in front of $\phi^2 R$, it would be possible to get a model with inflation. However, this would break the conformal invariance symmetry of the original model \rf{toy}  explicitly.

In the models of single component real fields $\chi$ and $\phi$ one cannot easily modify the coefficient in front of the term  ${\phi^2} R$ while maintaining conformal invariance of the original model. In the next section we will show that one can do it in a model of two complex scalar fields, along the lines of the superconformal models studied in \cite{Ferrara:2010in,Ferrara:2010yw}. As always in supersymmetry, the two complex scalar fields will  form a \K\, manifold. When such a \K\, manifold is flat, one finds the expected conformal coupling. However, when the \K\, manifold is curved, the action remains conformal invariant, as long as the conformon is not fixed. The curvature of such a  \K\, manifold is associated with the parameter $\Delta$ which will play an important role in our model. The deviation of $|\Delta|$ from its critical value
of $|\Delta|_{\rm cr}= 1/6$  will provide a possibility for the quartic inflaton potential to agree with the experiment, since 
$\xi =|\Delta| - |\Delta|_{\rm cr}$.

\section{A Bosonic Conformal $\Delta$-Model:  Simplified  Superconformal  Model}\label{scmatter}

Our starting point is  one of the superconformal models discussed in  \cite{Ferrara:2010yw,Ferrara:2010in}. However,  we simplify it here so that the model has only a local conformal symmetry \footnote{ The proper superconformal model has in addition to local Weyl symmetry also a local $U(1)\, \mr$ symmetry,  local special conformal symmetry,  local supersymmetry and local special supersymmetry. The bosonic part of the superconformal action preserves all bosonic symmetries.}. 

We do not require local supersymmetry and local special supersymmetry, when fermions are added, we do not require a local $U(1)\, \mr$ symmetry and a local special conformal symmetry for the bosonic action. Therefore we do not have to construct the potential from the superpotential and we do not need the gauge fields for the local $U(1)\, \mr$ symmetry. Both of these facts  
 lead to a significant simplification of the conformal model, we introduce below,  comparative to the superconformal models
 in  \cite{Ferrara:2010yw,Ferrara:2010in}.

We consider two complex scalars, the conformon $X^0$   and  the inflaton $X^1$. The relevant $\Delta$-deformed CSS action with local conformal symmetry is the following
\be
\mathcal{L}_{\rm c}= \sqrt{-g}\Big [  -\ft16\Big (-\left| X^{0}\right| ^{2}+\left|
X^{1 }\right| ^{2}\Big ) R +  {\Delta\over 2} |X^0|^2  \left[ \left(\frac{X^{1}}{X^{0}}\right)^{2}+\left(\frac{\bar X^{\bar 1}
}{\bar X^{\bar 0}}\right)^{2}\right]  R
-G_{I\bar J}\partial^\mu X^I\,\partial_\mu \bar X^{\bar J}-V(X, \bar X) \Big ] \, ,
 \label{SimplescGrav}
\end{equation}
\noindent where $I, J=0,1$ and $\bar I, \bar J= \bar 0, \bar 1 $.  The real \K\, potential of the embedding manifold  is given by the following function
\begin{equation}
\mathcal{N}\left( X,\bar{X}\right) =-\left| X^{0}\right| ^{2}+\left|
X^{1 }\right| ^{2}- 3\, \Delta  |X^0|^2  \left[ \left(\frac{X^{1}}{X^{0}}\right)^{2}+\left(\frac{\bar X^{\bar 1}
}{\bar X^{\bar 0}}\right)^{2}\right]   .
\label{N}\end{equation}
The action can be given in the form
\be
\mathcal{L}_{\rm c}= \sqrt{-g}\Big [  -\ft16{\cal N} \left( X,\bar{X}\right) R -G_{I\bar J}\partial^\mu X^I\,\partial_\mu \bar X^{\bar J}-V \Big ] \, .
 \label{scGrav}
\end{equation}
Here the $\Delta$-deformed kinetic terms  for scalars are 
\begin{equation}
G_{I\bar{J}}\equiv \frac{\partial ^{2}\mathcal{N}}{\partial
X^{I}\partial \bar{X}^{\bar{J}}}=\left(
\begin{array}{cc}
G_{0\bar{0}} & G_{0\bar{1 }} \\
G_{1 \bar{0}} & G_{1 \bar{1 }} 
\end{array}
\right) \ ,
\label{metric}\end{equation} 
\begin{eqnarray}
\hskip -0.5 cm G_{0\bar{0}} =-1+3\Delta    \left[ \left(\frac{X^{1}}{X^{0}}\right)^{2}+\left(\frac{\bar X^{\bar 1}
}{\bar X^{\bar 0}}\right)^{2}\right]  \, ,
~
G_{0\bar{1 }} =-6\, \Delta \, \frac{\bar{X}^{\bar{1 }}}{\bar{X}^{\bar{0}}%
}\, , ~
G_{1 \bar{0}} =-6\, \Delta \,  \frac{X^{1 }}{X^{0}}\, , ~ 
G_{1 \bar{1 }} =1.
\label{metric1}\end{eqnarray}
The action \rf{SimplescGrav} is invariant under local conformal transformations
\be
(X^I)'= e^{\sigma(x)} X^I, \qquad (\bar {X}^{\bar{J}})'= e^{\sigma(x)}  \bar {X}^{\bar{J}}\, , \qquad g_{\mu\nu}'= e^{-2\sigma(x)}g_{\mu\nu}
\label{conf2}\ee
{\it under condition that the potential $V(X, {\bar X})$ is homogeneous and second degree in both $X$ and $ {\bar X}$.} For example, one can take
\be
V= \lambda (X^1 \bar X^{\bar 1} )^2 \ ,
\label{pot}\ee
but one can consider other potentials as well.

The two complex scalars $X^I, \bar {X}^{\bar{I}}$ form a K{\"a}hler manifold with
metric in eq. \rf{metric}, and connection and curvature given,  by
\be \Gamma^I_{JK}= G^{I\bar L}{\cal N}_{JK\bar L}\ ,
\qquad 
  R_{I\bar K J\bar L}= {\cal N}_{IJ\bar K\bar L}-{\cal N}_{IJ\bar M}G^{M\bar M}{\cal N}_{M\bar K\bar L} \ ,
  \label{curv}\ee
where \be
{\cal N}_{JK\bar L}=\frac{\partial ^{3}\mathcal{N}}{\partial
X^{J} \partial
X^{K}\partial \bar{X}^{\bar{L}}} \, , \qquad {\cal N}_{IJ\bar K\bar L}= \frac{\partial ^{4}\mathcal{N}}{\partial
X^{I} \partial
X^{J}\partial \bar{X}^{\bar{K}} \partial \bar{X}^{\bar{L}}}  \ .
\label{curv1}\ee
In presence of $\Delta$ the connection and the curvature of the K{\"a}hler manifold do not vanish, at $\Delta=0$ the 
K{\"a}hler manifold is flat, the connection and curvature vanish and we recover an undeformed CSS model:
\be
\mathcal{L}_{\rm sc}= \sqrt{-g}\Big [  -\ft16\Big (-\left| X^{0}\right| ^{2}+\left|
X^{1 }\right| ^{2}\Big ) R 
-\delta _{I\bar J}\partial^\mu X^I\,\partial_\mu \bar X^{\bar J}-V \Big ] \, ,
 \label{SimplescGrav1}
\end{equation}
It coincides with the toy model  (2.1) for real fields $X^0=\chi/\sqrt 2 $ and $X^1=\phi/\sqrt 2 $ for $V= \lambda (X^1 \bar X^{\bar 1} )^2$.

\subsection{Enhanced symmetries of the \K\, potential of the embedding manifold}

Here we replace $\Delta $ by $\pm  ( \xi+1/6) $
\begin{equation}
\mathcal{N}\left( X,\bar{X}\right)_{\pm} =-\left| X^{0}\right| ^{2}+\left|
X^{1 }\right| ^{2}\mp  3 \Big ( \xi+{1\over 6}\Big )  |X^0|^2  \left[ \left(\frac{X^{1}}{X^{0}}\right)^{2}+\left(\frac{\bar X^{\bar 1}
}{\bar X^{\bar 0}}\right)^{2}\right]  .
\label{N1}\end{equation}
For positive $\Delta$ we have a case $\mathcal{N}\left( X,\bar{X}\right)_{+}$, for negative $\Delta$ we have a case $\mathcal{N}\left( X,\bar{X}\right)_{-}$.

At $\xi=-1/6$ ($\Delta=0$)  the \K\, potential has an enhanced $SU(1,1)$ symmetry. 
\begin{equation}
\mathcal{N}\left( X,\bar{X}\right)_{\pm} =-\left| X^{0}\right| ^{2}+\left|
X^{1 }\right| ^{2}.
\label{N11}\end{equation}
The second symmetry becomes manifest  at $\xi=0$ ($\Delta = \Delta_{\rm cr}= \pm 1/6$). In this case one can represent $\mathcal{N}\left( X,\bar{X}\right)_{\pm}$ as follows:
\begin{equation}
\mathcal{N}\left( X,\bar{X}\right)_{\pm} =
-\left | X^{0}\right| ^{2} \mp {1\over 2}  \left | X^{0}\right| ^{2}\left({X^1\over X^0} \mp {\bar X^{\bar 1}\over \bar X^{\bar 0}} \right)^2 .
\label{N2}\end{equation}
For $\Delta= +1/6$, this potential is invariant under the transformation with a real parameter $\Lambda$,
\bea\label{sym1}
X^1 \rightarrow  X^1 + \Lambda X^0\, , \qquad 
\bar X^1 \rightarrow  \bar X^1 +   \Lambda \bar X^0\, , \qquad \Lambda=  \bar \Lambda\, , 
\eea
For $\Delta = -1/6$, the potential in invariant under the transformation with an imaginary $\Lambda$,
\bea\label{sym2}
X^1 \rightarrow  X^1 +  \Lambda X^0\, , \qquad 
\bar X^1 \rightarrow  \bar X^1 +  \Lambda \bar X^0\ , \qquad \Lambda= - \bar \Lambda\,  \ .
\eea
This transformation mixes the inflaton $X^1$ with the conformon $X^0$. In terms of the homogeneous coordinate $z= X^1/X^0$, for $\Delta= +1/6$ this symmetry is a shift symmetry in the real direction 
\be
z\rightarrow z+\Lambda\, ,\qquad \bar z\rightarrow  \bar z + \Lambda\ ,  \qquad \Lambda=  \bar \Lambda\, ,
\ee
and for $\Delta= -1/6$ it is symmetry with respect to shift in the imaginary direction
\be
z\rightarrow z+\Lambda\, ,\qquad \bar z\rightarrow  \bar z + \Lambda \  , \qquad \Lambda= - \bar \Lambda\, .
\ee
For supergravity applications, this means that for $\Delta= +1/6$ the \K\ potential does not depend on the real part of the field $\Phi$, which can be  identified with the inflaton field in this context. Meanwhile for $\Delta= -1/6$ the \K\ potential does not depend on the imaginary part of the field $\Phi$, which can be identified with the inflaton field for $\Delta= -1/6$.

To summarize, the embedding \K\, potential  as a function of $\Delta=\pm  ( \xi+1/6)$ has two critical points with enhanced symmetry 
\begin{itemize}
  \item maximal enhanced symmetry,  undeformed CSS model \,  $\Delta=0,  \hskip 0.5 cm  \xi_{\rm re}= -1/6, \hskip 0.5 cm \xi_{\rm im}= -1/6$ 
  \item  double critical point of enhanced symmetry,  \, \, $\Delta=\Delta_{\rm cr}= \pm 1/6\, \, \hskip 0.3 cm  \xi_{\rm re/im}= 0, \hskip 0.4 cm  \xi_{\rm im/re}= -1/3$
\end{itemize}
It is the second case with $\Delta=\Delta_{\rm cr}= \pm 1/6, \hskip 0.5 cm \xi_{\rm re/im}= 0$, the case of a `critically deformed CSS model'  allows a nice interpretation of the Planck2013 data presented in Fig. 1, assuming that we can use
  $\Delta\approx \pm 1/6$ for the $\xi\approx 0$ for the inflaton.

\subsection{Spontaneous breaking of conformal symmetry}
Now we proceed with the analysis of the equations of motion of the locally conformal model \rf{SimplescGrav} with  the potential 
\rf{pot}. Note that this locally conformal model does not have a single dimensionful parameter.
To solve equations of motion of the conformal model \rf{SimplescGrav} we may use an ansatz for the conformon
\be
X^0=\bar X^{\bar 0}= \sqrt 3 M_P\, .
\label{Pl}\ee
This ansatz may be  associated either with the gauge-fixing of the local Weyl-${\mathcal{R}}$-symmetry of the superconformal action, or viewed as a part of a solution of all field equations following from the superconformal action. In this latter case, one may think of the origin of the Planck mass $M_P $ in this model as a spontaneous breaking of Weyl symmetry, via solution of equations of motion, since $M_P$ does not appear in the original locally conformal action  \rf{SimplescGrav}.
With the ansatz \rf{Pl} the action simplifies and becomes a part of the supergravity action
\be
\sqrt{-g}^{-1}\mathcal{L}_{\rm sg}=   \ft12 M_P^2 R - \ft16 
X^{1 }\bar{X}^1  R\ +  {\Delta\over 2}   \left( (X^{1
})^2+(\bar{X}^{\bar{1 }})^2\right) R
-\partial^\mu X^1\,\partial_\mu \bar X^{\bar 1}-\lambda (X^1 \bar X^{\bar 1} )^2  \, .
 \label{SimplescGravJ}
\end{equation}
Note that the only dependence of the model on $M_P$ enters in the Einstein action $\ft12 M_P^2 R$, the rest of the action does not depend on $M_P$.

We see that the coupling $- \ft16 X^{1 }\bar{X}^1  R$ is still  conformal, but the holomorphic and anti-holomorphic couplings
${\Delta\over 2} \left( (X^{1
})^2+(\bar{X}^{\bar{1 }})^2\right) R$ will change coupling with gravity for the real and imaginary part of $X^1 = {1\over \sqrt 2} (\varphi_1 + i  \varphi_2)$:
\be
\sqrt{-g}^{-1}\mathcal{L}_{\rm sg}=   \ft12 M_P^2 R -  \ft12 \Big ({1\over 6} - \Delta \Big) \varphi_{1}^{2} R  - \ft12 \Big({1\over 6}+ \Delta \Big)\varphi_{2}^{2}R 
-{\ft12} \partial^\mu \varphi_1\,\partial_\mu \varphi_1 - {\ft12} \partial^\mu \varphi_2\,\partial_\mu \varphi_2-  {\ft\lambda 4}  (\varphi_1^2 + \varphi_2^2)^2  \, .
 \label{SimplescGravJ1}
\end{equation}
The model has symmetry
\be
\Delta \rightarrow -\Delta\, , \qquad \varphi_1 \rightarrow \varphi_2 \, , \qquad \varphi_2 \rightarrow \varphi_1 \ .
\ee
 We make a choice $\Delta \geq 1/6 +\xi>0$ where the model has a minimum at $ \varphi_2=0$ and it is reduced to
\be
\sqrt{-g}^{-1}\mathcal{L}_{\xi}=   \ft12 M_P^2 R - {1\over 12} 
\, \varphi^2   \, R\ + {\Delta\over 2}\, \varphi^2  \, R
-{1\over 2} \partial^\mu \varphi\,\partial_\mu \varphi -  {\lambda\over 4}  \varphi^4  \, ,
 \label{SimplescGravJ2}
\end{equation}
where  $\varphi_1\equiv \varphi $. 
Thus, starting with spontaneously broken conformal symmetry, we have reproduced  the model ${\lambda\over 4} \phi^4 - {\xi\over 2}\phi^2 R$ studied recently in \cite{Okada:2010jf,Bezrukov:2013fca}, as well as the Higgs inflation models \cite{Salopek:1988qh,Sha-1,Einhorn:2009bh,Ferrara:2010yw,Lee:2010hj,Ferrara:2010in} where $\xi$ is a parameter of a non-minimal coupling to gravity. The difference is that in all these models, but \cite{Ferrara:2010yw,Ferrara:2010in}
 conformal symmetry was absent (i.e. broken explicitly).  The new interpretation of the parameter $\xi$ in our superconformal model with the embedding \K\, potential \rf{N1} is that at $\xi=0$ the \K\, potential has an enhanced symmetry \rf{sym1},  \rf{sym2}
 between the inflaton and a conformon.

\subsection{Explicit Proof of  Conformal Symmetry of the Action with Arbitrary $\Delta$}

Here we would like to explain why the action \rf{scGrav} has a local conformal symmetry for an arbitrary  parameter $\Delta$. This is rather surprising from the point of view of the models with scalars  conformally coupled to gravity, which were studied in the cosmological literature in the past. This action has a local conformal symmetry \rf{conf2}.
The \K\, potential of the embedding manifold\footnote{It is interesting that the \K\, potential of the embedding manifold does not allow \K\, transformations ${\cal N} (X,\bar X) \rightarrow {\cal N} (X,\bar X) + f(X) + \bar f(\bar X)$.}
${\cal N} (X,\bar X)$ is required to have the following properties \cite{Kallosh:2000ve},  \cite{Ferrara:2010in,Ferrara:2010yw,Freedman:2012zz}.
\be
{\cal N} (X,\bar X) = X^I {\cal N} _I= \bar X^{\bar J} {\cal N}_{\bar J}= X^I {\cal N}_{I\bar J}  \bar X^{\bar J} \ , \qquad {\cal N}_{I}\equiv \frac{\partial \mathcal{N}}{\partial
X^{I} } \ ,
\label{1}\ee
\be
{\cal N}_{IK\bar L} X^I= {\cal N}_{\bar I K\bar L} \bar X^{\bar I}=0 \ , \qquad {\cal N}_{IK\bar L} \bar X^L= {\cal N}_{IK} \ .
\label{2}\ee
The space-time curvature scalar in four dimensions  transforms as follows under local conformal transformations of the metric
\cite{Eisenhart}
\be
R' = (g^{\mu\nu})' R_{\mu\nu}'= e^{2\sigma} [ R - 6 g^{\mu\nu} \partial_\mu  \sigma \partial_\nu \sigma + 6 D^\mu D_\mu \sigma ] \ .
\label{Rvar}\ee
where $D_\mu$ implies the general covariant derivative.
 The infinitesimal form of the local conformal transformations is
 \be
\delta X^I= \sigma(x) X^I, \qquad \delta \bar {X}^{\bar{J}}= \sigma(x)  \bar {X}^{\bar{J}}\, , \qquad \delta g_{\mu\nu}= -2\sigma(x)g_{\mu\nu} \ .
\label{confInf}\ee
Therefore in the variation of the action we only need to keep the terms of the first order in $\sigma(x)$. Thus
\be
\delta R =  2\sigma  R  + 6 D^\mu \partial _\mu \sigma  \ .
\ee
We have to deal with terms depending on derivatives of $\sigma$ since all terms without derivatives on $\sigma$ clearly cancel. Thus we have to look at 
\be
- \sqrt{-g} \, {\cal N} (X,\bar X ) \, D^\mu \partial _\mu \sigma 
\label{R}\ee
from the first term in \rf{scGrav} and at
\be
-  \sqrt{-g} \,  G_{I\bar J}[ \partial ^\mu(\sigma X^I)\,\partial_\mu \bar X^{\bar J} +  \partial^\mu X^I\,\partial_\mu (\sigma \bar X^{\bar J})]
\ee
from the second term in \rf{scGrav}.
From this term we extract only a term with derivatives on $\sigma(x)$
\be
 -  \sqrt{-g} \,  G_{I\bar J} \,\partial_\mu (   \, X^I  \bar X^{\bar J} ) \partial^\mu \sigma \ .
\ee
We may rewrite this term, up to total derivatives, as follows
\be
  \sqrt{-g} [  G_{I\bar J} \,    \, X^I  \bar X^{\bar J}  D^\mu \partial^\mu \sigma +    \partial^\mu G_{I\bar J} \,    \, X^I  \bar X^{\bar J}   \partial^\mu \sigma] \ .
\label{kin}\ee
The first term in this expression, using \rf{1},  can be given in the form
\be
  \sqrt{-g}  \,  {\cal N} (X,\bar X ) \,  D^\mu \partial^\mu \sigma \ ,
\label{kin1}\ee
and it cancels the contribution from the curvature term \rf{R}. The remaining term from the variation of the kinetic energy of scalars is
\be
    \sqrt{-g} \, (\partial_\mu \, G_{I\bar J}) \,    \, X^I  \bar X^{\bar J}   \partial^\mu \sigma \ .
\ee
To prove that it vanishes we need to use the fact that $G_{I\bar J}= {\cal N}_{I\bar J}$ and note that
 due to  \rf{2}
\be
(\partial_\mu \, {\cal N}_{I\bar J} )\,    \, X^I  \bar X^{\bar J}= \partial_\mu X^K {\cal N}_{K I\bar J}  \, X^I  \bar X^{\bar J}+ 
\partial_\mu \bar X^{\bar K} {\cal N}_{\bar K  I\bar J} \, X^I  \bar X^{\bar J}=0 \ .
\ee
This completes the proof that all terms in the variation of the action \rf{scGrav} depending on derivatives of $\sigma(x)$, for small $\sigma(x)$ cancel. One can also check that terms with $6 g^{\mu\nu} \partial_\mu  \sigma \partial_\nu \sigma$ from the $R$-variation in \rf{Rvar} cancel from scalar kinetic terms with account of the relation \rf{1}.

For $x$-independent $\sigma$ the symmetry of the action \rf{scGrav} is trivial since ${\cal N}(X, \bar X)$ is homogeneous of first degree both in $X$ and in $\bar X$. Therefore each  term in \rf{scGrav} is scale-invariant for constant $\sigma$. However, for local conformal transformations \rf{confInf} the symmetry of the action is valid due to cancellation of the contributions from the curvature terms and scalar kinetic term. The precise deviation of kinetic terms from the canonical ones in $-{\cal N}_{I\bar J}\partial ^\mu X^I\,\partial _\mu \bar X^{\bar J}$ due to $\Delta$ compensates the variation of the curvature terms due to $\Delta$ terms in $-{1\over 6} \, {\cal N}\, R$.

It remains to confirm that  our model \rf{N} indeed, satisfies the crucial properties of the \K\, potential of the embedding manifold, given in eqs.  \rf{1}, \rf{2}. One can either check them directly, or notice that these properties follow from the  property of the \K\, potential to be homogeneous first degree in $X$ and in $\bar X$:
\be
{\cal N}(a X, \bar a \bar X)=a \, \bar a \, {\cal N}( X,  \bar X) \ ,
\ee
where $a$ is an arbitrary complex number.

\section{ $\Delta$-Deformed Canonical Superconformal Supergravity Model}\label{SUGRA}

\subsection{The role of  Goldstino}
The major difference between the bosonic conformal model above and the superconformal model, which we will describe here,  is that in addition to a conformon $X^0$ and inflaton $X^1=\Phi$ we need one more superfield $X^2=S$, a Goldstino.

The complete
 gravity-scalar part\footnote{A generic  superconformal theory\cite{Kallosh:2000ve,Ferrara:2010in,Ferrara:2010yw} underlying  all possible N=1 supergravities  contains  superconformal-invariant terms describing gravity and chiral multiplets and gauge multiplets. The gauge part of the superconformal model \cite{Ferrara:2010in,Ferrara:2010yw}   was recently used  to describe the superconformal D-term inflation in \cite{Buchmuller:2012ex}.} of the $SU(2,2|1)$ invariant superconformal action \cite{Kallosh:2000ve,Ferrara:2010in,Ferrara:2010yw}  has a gravity part, kinetic terms for scalars and a potential.  Ignoring fermions, it is given by
\be
{1\over \sqrt{-g}}\mathcal{L}_{\rm sc}^{\rm scalar-grav}=-\ft16{\cal N} (X,\bar X)R
-G_{I\bar J}{\cal D}^\mu X^I\,{\cal D}_\mu \bar X^{\bar J}-G^{I\bar J}{\cal W}_I \bar{{\cal W}}_{\bar J} \, , \qquad I, \bar I = 0,1,2.
 \label{scGrav11}
\end{equation}
Here, the $U(1)\, \mr$ symmetry covariant derivative   is
\begin{eqnarray}
{\cal D}_\mu X^I & = & \partial_\mu X^I -\rmi A_\mu
X^I   \, , \label{covderconf1}
\end{eqnarray}
where  the local special conformal symmetry has been already gauge-fixed via the $b_\mu=0$ condition, and $A_\mu$ is a gauge field of the 
the  local $U(1)\, \mr$ symmetry.
As above, 
$
  G_{I\bar J}=\partial_I \partial_{\bar J} {\cal N} \equiv {\partial  \mathcal{N}(X, \bar X)\over \partial X^I \partial \bar X^{\bar J}}
\label{LargeK}  $, 
however, the F-term potential 
\be
V_F= G_{I\bar J}  F^I \bar F^{\bar J}=G^{I\bar J}{\cal W}_I \bar{ {\cal W}}_{\bar J} 
\ee
must be constructed from the superpotential since it originates from the solutions for the auxiliary fields for the chiral multiplets.  Each superfield $X^I$ includes a scalar, a spinor and an auxiliary field $F^I$ whose value is defined by the derivatives of the superpotential $F^I= G^{I\bar J} \bar {\cal W}_{\bar J}$,
\be
{\cal W}_I\equiv {\partial {\cal W}\over \partial X^I}\ , \qquad  \bar{{\cal W}}_{\bar J}\equiv  {\partial \bar{{\cal W}}\over \partial \bar X^{\bar J}} \ .
\ee
In the absence of fermions  the action (\ref{scGrav11}) has a  local conformal and local $U(1)$ ${\mathcal{R}}$-symmetry, which is part of the superconformal $SU(2,2|1)$ 
symmetry. This means that the action (\ref{scGrav11}) is invariant under the following transformations:
\be
(X^I)'= \rme^{\sigma(x)+\rmi\Lambda(x)}  X^I\, , \qquad (\bar X^I)'= \rme^{\sigma(x)-\rmi\Lambda(x)}  \bar X^I  \ ,
\ee
\be
g_{\mu\nu} '= \rme^{-2\sigma(x)} g_{\mu\nu}\, , \qquad A_\mu '= A_\mu +\partial _\mu \Lambda(x)
 \ .
\label{dil}\ee
The  chiral multiplets in our model $X^I$, include the compensator field $X^0$, the inflaton $X^1=\Phi $ and the Goldstino  superfield $X^2=S$
\be
X^I= (X^0, X^1=\Phi, X^2=S) \ .
\ee
 The role of the conformon field $X^0$  in the action is to support an unbroken  superconformal symmetry. During inflation the conformon breaks spontaneously the Weyl symmetry and the $\mathcal{R}$ $ U(1)$ symmetry when equations of motion of the superconformal theory  are solved with an ansatz
$
X^0=\bar X^{\bar 0}= \sqrt 3 M_P$. The inflaton and Goldstino together break spontaneously the local supersymmetry when the solution of equations of motion has a non-vanishing auxiliary field $F_{S}(\Phi)\equiv {\partial {\cal W}\over \partial {S}}\neq 0$.  The potential during inflation depends on the inflaton as 
\be
V(\Phi)= |F_{S}(\Phi)|^2= \Big |{\partial {\cal W}(\Phi, S)\over \partial {S}}\Big |^2
\ee 
{\it under condition that the superpotential is linear in} $S$. We will find that during inflation $S$ direction is indeed a Goldstino direction since only in $S$ direction the auxiliary field $F_S$ is not vanishing \cite{Kallosh:2010ug},
\be
F_0= {\partial {\cal W}\over \partial X^0}=0\, ,\qquad F_\Phi= {\partial {\cal W}(S, \Phi)\over \partial \Phi}=0
\, ,\qquad F_S= {\partial {\cal W}(S, \Phi) \over \partial S} \neq 0 \ .
\ee
The reason for this is that in our class of models the scalar field $S$ vanishes during inflation.  Therefore the bosonic conformal model of the previous section provides a simplified version of the full superconformal model.

\subsection{Superconformal generalization of the model ${\lambda\over 4} \phi^4 - {\xi\over 2}\phi^2 R$}
In Appendix we present a CSS model which even after local conformal symmetry is spontaneously broken via the Einstein term, still provides a conformal coupling of the inflaton to gravity. However, this model does not support inflation, unless we break the superconformal symmetry of the inflaton.

Fortunately, one can preserve important advantages of canonical superconformal supergravity models and describe inflation if one supplement the K{\"a}hler potential and the superpotential by certain terms some of which vanish on inflationary trajectory. Here we present the simplest model of this type \cite{Ferrara:2010in}:

\noindent 1.  We deform a
flat $SU(1,2)$ K{\"a}hler manifold of a canonical model \rf{calNminimal} for our 3 superfields, $ X^I= (X^0, \Phi, S)$ and $ \bar X^{\bar I}= (\bar X^{\bar 0}, \bar \Phi, \bar S)$
   as follows
\be
\mathcal{N}(X,\bar X)= -|X^0|^2 + |\Phi|^2 + |S|^2 - 3 \Delta |X^0|^2  \left[ \left(\frac{\Phi
^2}{X^{0}}\right)^{2}+\left(\frac{\bar\Phi
^2}{\bar X^{\bar 0}}\right)^{2}\right] - 3 \zeta {(S\bar S)^2\over |X^0|^2}\,  \,.
\label{calNminimal2}
 \ee
 This means that  the kinetic terms $G_{I\bar J}={\cal N}_{I\bar J}$ for all multiplets  in general are not canonical, except 
$
 \eta_{\Phi \bar \Phi }=1  $ which remains canonical since the $\Delta$-terms have a holomorphic and an anti-holomorphic dependence on $\Phi$. Note that  $\Delta= \pm (\xi+1/6)$ where $\xi$ is  an essential parameter relevant for the physics of the model. The $\zeta$ term is required for the stability of the $S=0$ extremum of the potential \cite{Lee:2010hj, Ferrara:2010in}. During inflation this term does not contribute to a cosmological evolution since it vanishes at $S=0$.

 \noindent 2. We keep the same  cubic superpotential  ${\cal W}(\Phi, S)
=  \sqrt \lambda \, \, S\,  \Phi^2$ as in the canonical model, and it is  $X^0$ independent. The  bosonic part of the $\xi$-model superconformal action is
 \begin{eqnarray}
 {1\over \sqrt{-g}}{ \mathcal{L}}_{\rm sc}^{\xi}&=&{1\over 6}\Big [|X^0|^2 - |\Phi|^2 - |S|^2 \pm 3(\, \xi + 1/6)\, |X^0|^2 \left[ \left(\frac{\Phi
^2}{X^{0}}\right)^{2}+\left(\frac{\bar\Phi
^2}{\bar X^{\bar 0}}\right)^{2}\right] - 3 \zeta {(S\bar S)^2\over |X^0|^2}\Big ] R \nonumber\\
\nonumber\\
&-&G_{I\bar J}{\cal D}^\mu X^I\,{\cal D}_\mu \bar X^{\bar J}+
 \lambda (  G^{S\bar S} |\Phi\bar \Phi|^2 + 4 G^{\Phi \bar \Phi}|S\Phi|^2)\,   \ .
 \label{delta} 
\end{eqnarray}
Now both  $\Phi$ as well as  $S$ do  interact with the conformon $X^0$ when $\xi \neq -1/6$ and $\zeta\neq 0$.

If we gauge-fix the conformon, $X^0=\bar X^{\bar 0}= \sqrt 3 M_P$,  which corresponds to a spontaneously broken superconformal symmetry, the gravity part will be the same as in \rf{grav1},  
$ {1\over 2} M_P^2( R
+ 6 A^\mu A_\mu) 
$ and will depend on $M_P$.
However, 
 the action of $\Phi$ and  $S$ will change. Instead of the canonical superconformal action \rf{mat} we find
 \begin{eqnarray}
&& {1\over \sqrt{-g}}{ \mathcal{L}}_{\rm mat}^{\xi}=- {1\over 6} \Big [ |\Phi|^2 +|S|^2       \pm 3(\xi+1/6)   ( \Phi
^2+\bar{\Phi }^2)   - \zeta  {(S\bar S)^2\over M_P^2} \Big]R  \nonumber\\
\nonumber\\
&& -{\cal D}^\mu \Phi\,{\cal D}_\mu \bar \Phi -G_{S\bar S} {\cal D}^\mu S\,{\cal D}_\mu \bar S
-\lambda (  G^{S\bar S} |\Phi\bar \Phi|^2 + 4 G^{\Phi \bar \Phi}|S\Phi|^2) \ .
 \label{matdelta}
\end{eqnarray}
In this form the  {\it maximal enhanced symmetry on the gravity coupling to matter} is manifest at
\be
\xi=-1/6\, ,  \qquad \zeta=0 \ ,
\ee
where we recover the CSS model
\be
{1\over \sqrt{-g}}{ \mathcal{L}}_{\rm coupling}^{\rm gravity/matter }= - {1\over 6} \Big [ |\Phi|^2 +|S|^2\Big]R \ .
\ee
We may rewrite the matter/gravity coupling  in the form which makes a {\it minimal  enhanced symmetry} manifest at 
\be
\Delta= \Delta_{\rm cr} = \pm 1/6\, , \qquad \xi_{\rm re/im}=0 \ ,
\ee
 \begin{eqnarray}
 {1\over \sqrt{-g}}{ \mathcal{L}}_{\rm coupling}^{\rm gravity/matter }= {1\over 6} \Big [ \pm 1/2 (\Phi \mp \bar \Phi)^2
\mp 3\xi  ( \Phi
^2+\bar{\Phi }^2) 
 -|S|^2         + \zeta  {(S\bar S)^2\over M_P^2} \Big]R  \ .
 \label{matdeltaw}
\end{eqnarray}
In this form at $\xi=0$ the symmetry is $\Phi\rightarrow \Phi+ \Lambda$ and $\bar \Phi \rightarrow \bar \Phi+ \bar \Lambda$ with $\Lambda = \bar\Lambda $ at the positive critical point $\Delta= \Delta_{\rm cr} =  1/6$ and $\Lambda = -\bar\Lambda $ at the negative critical point $\Delta= \Delta_{\rm cr} = - 1/6$.
 This symmetry is broken at $\xi\neq 0$.

The cosmological evolution of this model during inflation was studied in earlier papers \cite{Ferrara:2010in,Lee:2010hj,Ferrara:2010yw,Kallosh:2010ug, Linde:2011nh} 
where it was concluded that there is a minimum of the potential for positive or negative  $\Delta$ at 
\be
S=\bar S= 0\, ,  \qquad \Phi=\pm \bar \Phi\, ,  \qquad A_\mu=0 \ .
\label{min}\ee
We are interested in small values of $\xi$ which make this model in agreement with Planck data starting with $\xi\geq  2\cdot 10^{-3}$. Already at $\xi=0.1$ the model reaches the regime of large non-minimal coupling $\xi $ and does not change significantly anymore when $\xi $ is increasing and the coupling $\lambda$ is adjusted. A stability in the $S$ direction for such small values of $\xi$ is easy to achieve with $\zeta\geq1/6$.

Note that the dependence on $M_P$ in \rf{matdelta} via the term with $\zeta$ and via $G^{S\bar S}$ and $G^{\Phi \bar \Phi}$ disappears at $S=0$ at the inflationary trajectory. In this sense   the model ${\lambda\over 4} \phi^4 - {\xi\over 2}\phi^2 R$  a minimal deformation of the  CSS model. We will discuss the case of more general models, where all other deviation from CSS lead to explicit dependence on $M_P$ in the inflaton action even during inflation.

As a result, the effective action for the inflaton at the minimum \rf{min} in the $\xi$-model  is 
 \begin{eqnarray}
 {1\over \sqrt{-g}} { \mathcal{L}}_{\rm mat}^{\xi} =-  {1\over 2} \xi \phi^2  R 
 -{1\over 2} \partial ^\mu \phi\,\partial _\mu \phi 
-{\lambda\over 4} \phi^4  \ ,
 \label{mindelta}
\end{eqnarray}
where the real or the imaginary part of $\Phi$ is related to a single  scalar $\phi$ by a factor of $1/ \sqrt 2$. Along the inflationary trajectory with $S=0$ the supergravity model discussed above coincides with the bosonic model discussed in Section \ref{scmatter}. 

Thus, once again, we reproduced the inflaton potential of the type  proposed in \cite{Salopek:1988qh,Sha-1,Einhorn:2009bh,Ferrara:2010yw,Lee:2010hj,Ferrara:2010in}. However, now this model became a part of supergravity, and it has an underlying superconformal symmetry, which is  broken spontaneously. The supergravity version of this model is described by the physical moduli space \K\, potential and superpotential in $M_P^2=1$ units in the Einstein frame with
\be
K_{\pm}= -3\log\Big ( 1-{1\over 3} |S|^2 \pm {1\over 6} (\Phi \mp \bar \Phi)^2
\mp \xi  ( \Phi
^2+\bar{\Phi }^2 
) 
- {\zeta \over 3 }(S\bar S)^2\Big )
\label{Kal}\ee
and
\be
W= \sqrt \lambda \, S \, \Phi^2 \ .
\label{sup}\ee
 If $\zeta\geq 1/6$  and $\xi > 2\cdot 10^{-3} $, the minimum of the potential with $\Phi= {1\over \sqrt 2}(\phi_1+i\phi_2)$ in $K_+$ case  is at  $S=\phi_2=0$ and in $K_-$ case  is at is $S=\phi_1=0$. Thus the coupling of a single real field $\phi$ in the
 action \rf{mindelta} is recovered from supergravity with \rf{Kal} and \rf{sup}. This supergravity originates 
  from the superconformal  model \rf{delta}  upon spontaneous breaking of a superconformal symmetry.

\section{Phenomenology} 

For completeness, we should give here values of various parameters describing inflation in this class of models. This is a simple task, since they coincide with the parameters for the model $\lambda\phi^{4}/4$ of a real scalar field, which can be found in \cite{Bezrukov:2013fca}. 
The quartic coupling of the inflaton is
\be
\lambda = (1+6\xi)   \ { {{3\pi^{2}\over 2} \Delta_{{\cal R}}^2}\,  ( 1+6\xi+8(N+1)\xi) \over (1+8(N+1)\xi)  (N+1)^3}  \ ,
\ee
where one can use the WMAP9 normalization \cite{Hinshaw:2012aka}
$
 \Delta_{{\cal R}}^2 \approx 2.46\times 10^{{-9}} 
$. $N$ here is the number of e-foldings of inflation.
Slow roll parameters are
\be
r= {16 \, (1+6\xi) \over (N+1) (1+ 8(N+1) \xi)}\, , \qquad 
n_s = 1 - {3 \, (1+6\xi)  + 8(N+1) (5+ 24 \xi) \xi  + 128 (N+1)^2 \xi ^2 \over (N+1) (1+ 8(N+1) \xi)^2} \ .
\ee

 In \cite{Bezrukov:2013fca} it was argued that the number of e-foldings  in this model is close to 60  practically independently of the detailed theory of reheating. Indeed, equation of state of the oscillating inflaton field in the theory $\lambda\phi^{4}$ approximately coincides with the equation of state of relativistic particles produced during reheating, so the post-inflationary universe expands as $\sqrt t$ both before reheating and shortly after it.  This would reduce the range of possible values of $n_{s}$ and $r$ in this model to the single red line in Fig. 1, corresponding to $N \approx 60$.

If we consider inflation in supergravity, then we have two different options to consider, which may lead to somewhat different value of N. 

First of all, the inflaton field may belong to the visible sector, as in the Higgs inflation. In this case, reheating temperature is very high \cite{GarciaBellido:2008ab,Bezrukov:2008ut}
and gravitinos are excessively produced \cite{Ferrara:2010in}. This is not necessarily a problem if gravitinos are superheavy, as is the case with some recent models of mini-split supersymmetry, see e.g. \cite{Dudas:2012wi}   and references therein. 

However, if the field $\phi$ is not the Higgs field of the standard model, then it may be preferable to have inflation with the field $\phi$ belonging to the hidden sector. In this case, most of the particles produced during reheating will also belong to the hidden sector. Decay of these particles to matter in the visible sector is Planck mass suppressed, so it occurs with a large time delay; the reheating temperature is much smaller, which helps to solve the gravitino problem. Therefore the hidden sector particles produced during reheating may become non-relativistic and temporarily dominate the energy-density of the universe. As a result, the required number of e-foldings will be somewhat smaller than 60, as shown in Fig. 1. This may allow one to distinguish between the theory where the inflaton belongs to the visible sector, such as the Higgs inflation scenario, and the theory based on supergravity where the inflaton may belong to the hidden sector. This is very similar to the difference between the predictions of perturbations \cite{Mukhanov:1981xt}
 in the Starobinsky model \cite{Starobinsky:1980te} and in the Higgs inflation \cite{Salopek:1988qh,Sha-1,Einhorn:2009bh,Ferrara:2010yw,Lee:2010hj,Ferrara:2010in}, as analyzed in  \cite{Bezrukov:2011gp}.

The difference between these two implementations can be studied observationally, even though it is not an easy task. For example,   models of this type with $\xi \gtrsim 1$ predict $n_{s}\approx 0.967$ for $N = 60$ and $n_{s}\approx 0.963$ for $N = 55$. Both of these values are well inside the dark blue area in Fig. 1, but the last one gives a slightly better fit to Planck data. In this respect, additional flexibility provided by the supergravity implementation of this scenario is quite welcome.

\section{Discussion}

The main goal of this paper was to explain basic principles of construction of a family of models which match observational predictions of the theory $\lambda\phi^{4}$ with non-minimal coupling to gravity $\xi$, while having an additional advantage of being a part of a theory with spontaneously broken conformal or superconformal invariance. The first of these two goals can be achieved in a theory of two complex fields, a conformon and an inflaton, as shown in Section \ref{scmatter}. The second goal is achieved in the superconformal model  described in  Section \ref{SUGRA}; a Goldstino superfield $S$ plays an important role there, in addition to a conformon and an inflaton. The bosonic part of the superconformal model  coincides with the simplified conformal model when the Goldstino vanishes at the minimum of its potential, and the inflaton potential obtained from the superpotential $|D_{S}W|^{2}$ is  quartic in the inflaton.

Both of these models have the same inflaton potential as the simple conformally non-invariant model of a single real scalar field ${\lambda\over 4} \phi^4  - {\xi\over 2} \phi^{2}R$. However, now this model is embedded into a theory with spontaneously broken conformal or superconformal invariance. As we have seen, the possibility to embed the theory ${\lambda\over 4} \phi^4  - {\xi\over 2} \phi^{2}R$ into a theory with spontaneously broken conformal invariance with $\xi \not = -1/6$ is quite nontrivial. It was necessary to embed the original model of the real scalar field into a model of a complex scalar field of a very special type, or embed it into a closely related superconformal theory. 

In the locally conformal or superconformal model underlying the ${\lambda\over 4} \phi^4  - {\xi\over 2} \phi^{2}R$ model there is a single parameter $\Delta$ signaling a deformation of  the canonical conformal (superconformal) theory
with flat \K\, manifold of the embedding space.
$\Delta$  is a measure of the non-flatness of this manifold.  
When $\Delta=0$ we recover the canonical conformal (superconformal) theory with the flat \K\, manifold presented in the Appendix A. This corresponds to a conformal coupling of the inflaton to gravity, $\xi=-1/6$. However, the locally conformal and  superconformal theory, described in Sections 3 and 4, respectively,  have an additional double critical point when 
\be 
 \Delta= \Delta_{\rm cr} =\pm 1/6\, , \qquad \xi_{\rm re/im}=0 \, , \qquad  \xi_{\rm im/re}=-1/3 \ .
\ee
 At this point the \K\, potential of an embedding manifold, being non-flat,   has an enhanced symmetry
\bea
X^1 \rightarrow  X^1 + \Lambda X^0\, , \qquad 
\bar X^1 \rightarrow  \bar X^1 +  \bar \Lambda \bar X^0\, , 
\label{symX}\eea
where $X^1$ is a complex inflaton and $X^0$ is a complex conformon and the symmetry is mixing
the inflaton $X^1$  with the conformon $X^0$. 
For example, for $\Delta_{\rm cr} = 1/6 $ and 
$\xi_{\rm re}=0$ the symmetry is with  $  \Lambda=  \bar \Lambda\, $. In such case a small deviation of $\xi_{\rm re}$
from zero together with stabilization of the imaginary part of the field to zero, lead to a ${\xi\over 2} \phi^2 R - \lambda \phi^4$ model where the inflaton is a real part of the complex scalar field.

Thus, in this paper we described in detail the superconformal version of the $\lambda\phi^{4}$ model, which appears to be a  minimal deformation of a  canonical superconformal supergravity model \cite{Ferrara:2010in}.   The inflationary model with one single parameter $\xi$ derived from a model with spontaneous breaking of the superconformal symmetry has distinctive observational predictions. At  $\xi=0$ there is an enhanced symmetry of the embedding \K\, potential shown in eq. \rf{symX}. This  model therefore  suggests an interesting example of the  technical  naturalness \cite{'tHooft:1979bh} where there is a dimensionless parameter $\xi$, such that when this parameter is zero,  some symmetry of a system is restored.\footnote{We are grateful to G. t'Hooft for a clarification of this concept.} The symmetry in our case is a particular symmetry between the inflaton and the conformon fields in the underlying superconformal theory. In supergravity, after the superconformal symmetry is broken spontaneously, this symmetry corresponds to a shift symmetry of the quadratic part of the \K\, potential, as one can see from \rf{Kal}.

At the critical point at $\xi=0$ this symmetry is unbroken, but the amplitude of tensor perturbations in the theory $\lambda\phi^{4}$ is too high; but when this symmetry is even very slightly broken,  $\xi\geq  2\cdot  10^{-3}$, the cosmological predictions of this theory are in perfect agreement with Planck2013 results,  as shown by the red stripe in Fig. 1.  

In the subsequent set of papers, we will discuss other versions of inflationary models based on conformal and superconformal symmetry, and on supergravity, including the ones described in \cite{Kallosh:2010ug}, \cite{Linde:2011nh},
  \cite{tobepubl}.
A distinguishing feature of the ${\xi\over 2} \phi^2 R - \lambda \phi^4$ model is that superconformal symmetry is spontaneously broken there only by terms proportional to $R$, see eq. \rf{mindelta}, which do not introduce any mass scale, do not affect  masses of elementary particles, and become vanishingly small at the present stage of cosmological evolution.


\section*{Acknowledgments}

We are grateful to our collaborators S. Ferrara, A. Marrani  and A. Van Proeyen  for collaboration on the earlier papers on which this project is based, and for the useful recent discussions.  We also would like to acknowledge the stimulating discussions  with   D. Freedman, J. Garcia-Bellido, J.J. Carrasco, L. Dixon, D. Green, S. Kachru,  W. Siegel, E. Silverstein, G. t'Hooft,  P. Townsend,  and A. Westphal.
This work  is supported by the SITP and by the 
NSF Grant No. 0756174, the work of RK is also supported by the Templeton Foundation Grant ``Frontiers of Quantum Gravity''.  


\appendix

\section{ Canonical Superconformal Supergravity Model}

The generic class of CSS models discovered in \cite{Ferrara:2010in} has  extremely simple properties in a Jordan frame: the kinetic terms are canonical, the potentials are  the same as in globally supersymmetric theories. Below we describe an example of a corresponding model for a conformon, inflaton and  goldstino superfields. 

\noindent 1.  We choose a
 {\it flat $SU(1,2)$ K{\"a}hler manifold for our 3 superfields   }
\be
\mathcal{N}(X,\bar X)= -|X^0|^2 + |\Phi|^2 + |S|^2\,  \,.
\label{calNminimal}
 \ee
 This means that  $G_{I\bar J}={\cal N}_{I\bar J}= \eta_{I\bar J} \,, G^{I\bar J}= \eta^{I\bar J}$ where $I=0, 1,2$
 , and
$
 \eta_{0\bar 0}=-1,  \, \eta_{\Phi \bar \Phi }=\eta_{S \bar S }=1 \,.
 $
 
 \noindent 2. We choose a cubic  $X^0$-independent superpotential: 
 \be
{\cal W}(\Phi, S)
=  \sqrt \lambda \, \, S\,  \Phi^2\, ,  \qquad {\cal W}_0\equiv {\partial {\cal W} \over \partial X^0}=0 \ .
\label{superpot}\ee
We impose  our 2 conditions above on  the superconformal action (\ref{scGrav11})  and find 
 \begin{eqnarray}
 {1\over \sqrt{-g}}{ \mathcal{L}}_{\rm sc}^{can}={1\over 6}(|X^0|^2 - |\Phi|^2 - |S|^2) R
+D^\mu X^0\,D_\mu \bar X^{\bar 0}-D^\mu \Phi\,D_\mu \bar \Phi -D^\mu S\,D_\mu \bar S
-\lambda (  |\Phi\bar \Phi|^2 + 4 |S\Phi|^2)\nonumber\\
 \label{can}
\end{eqnarray}
A special feature of this model is that $\Phi$ and $S$ do not interact with the conformon $X^0$, their actions and equations of motion are conformally invariant. The total action consists of the gravity part 
\begin{eqnarray}
 {1\over \sqrt{-g}}{ \mathcal{L}}_{\rm sc}^{\rm grav}={1\over 6}|X^0|^2 R
+D^\mu X^0\,D_\mu \bar X^{\bar 0} \ ,
 \label{grav}
\end{eqnarray}
and the matter part
\begin{eqnarray}
 {1\over \sqrt{-g}}{ \mathcal{L}}_{\rm sc}^{\rm mat}=- {1\over 6}(  |\Phi|^2 + |S|^2) R
-D^\mu \Phi\,D_\mu \bar \Phi -D^\mu S\,D_\mu \bar S
-\lambda (  |\Phi\bar \Phi|^2 + 4 |S\Phi|^2) \ .\nonumber\\
 \label{mat}
\end{eqnarray}
Even if we gauge-fix the  local conformal and $U(1)\, \mr$ symmetry by making a choice $X^0=\bar X^{\bar 0}= \sqrt 3 M_P$, only the gravity part will change and become \begin{eqnarray}
 {1\over \sqrt{-g}}{ \mathcal{L}}_{\rm gauge-fixed}^{\rm grav}={1\over 2} M_P^2( R
+ 6 A^\mu A_\mu)  \ .
 \label{grav1}
\end{eqnarray}
However, the matter part \rf{mat} will still be superconformal. This means, in particular, that the energy-momentum tensor of $\Phi$ and $S$ is traceless. Therefore such models do not support inflation and we have to break the $SU(1,2)$ symmetry of the K{\"a}hler manifold for our 3 superfields  by introducing a parameter $\Delta$ which makes it non-flat.

\end{document}